\documentclass[aps,prl,reprint,groupedaddress,showpacs]{revtex4-1}

\usepackage{amsmath,amssymb}
\usepackage{mathbbol}
\usepackage{mathrsfs}
\def\bq{\pmb{q}}
\def\bp{\pmb{p}}

\def\bA{\pmb{A}}
\def\e{\mathrm{e}}
\def\d{\mathrm{d}}
\def\onehalf{{\textstyle\frac{1}{2}}}
\def\pfrac#1#2{\frac{\partial #1}{\partial #2}}
\def\dfrac#1#2{\frac{\d #1}{\d #2}}
\def\RB{\mathbb{R}}
\def\MB{\mathbb{M}}

\def\Hv{e}

\begin{document}


\title{Relativistic generalization of Feynman's path integral\\
on the basis of extended Lagrangians}


\author{J\"urgen Struckmeier}
\email[]{struckmeier@fias.uni-frankfurt.de}
\affiliation{Frankfurt Institute for Advanced Studies (FIAS),
Frankfurt am Main, Germany}

\date{\today}

\begin{abstract}
In the extended Lagrange formalism of classical point dynamics, the system's dynamics is parametrized along a system evolution parameter $s$, and the physical time $t$ is treated as a \emph{dependent} variable $t(s)$ on equal footing with all other configuration space variables $q^{i}(s)$.
In the action principle, the conventional classical action $L\,dt$ is then replaced by the generalized action $L_{\e}ds$.
A direct reparametrization of the conventional action yields an extended Lagrangian that is homogeneous of degree one in the velocities.
A regular extended Lagrangian that is not homogeneous of degree one can also reproduce the conventional dynamics on a suitably selected invariant hypersurface.
With this representation, the physical motion takes place on a hypersurface within a \emph{symplectic extended} phase space of even dimension.
The formal similarity of conventional and extended Lagrange formalisms provides a direct connection with the proper-time path integral approach to \emph{relativistic} quantum physics.
To provide an example, the non-homogeneous extended Lagrangian $L_{\e}$ of a classical relativistic spinless point particle in an external electromagnetic field will be presented.
This extended Lagrangian is a quadratic function in the velocities, and its selected zero-energy hypersurface is the classical mass shell.
With this $L_{\e}$, the generalized path integral first yields an evolution equation in $s$.
Imposing the corresponding quantum mass-shell condition then yields the Klein-Gordon equation as the physical wave equation.
This establishes the relation between the classical extended Lagrangian description and proper-time quantization for a charged spin-zero particle.
\end{abstract}


\maketitle

The conventional formulation of the principle of stationary action
is based on the action functional $S[\bq(t)]$, defined by
\begin{equation}\label{principle0}
S[\bq(t)]=\int_{t_{a}}^{t_{b}}L\left(\bq,\dfrac{\bq}{t},t\right)\d t,
\end{equation}
with $L(\bq,\dot{\bq},t)$ denoting the system's conventional Lagrangian,
and $\bq(t)=(q^{1}(t),\ldots,q^{n}(t))$ the vector of configuration
space variables as a function of time.
In this formulation, the independent variable $t$ is the coordinate time in a chosen reference frame.
A space-time formulation of the stationary action principle~(\ref{principle0})
is accomplished by treating
the time $t(s)=q^{0}(s)/c\,$ --- like the vector $\bq(s)$ of configuration
space variables --- as a \emph{dependent} variable of a newly
introduced independent evolution parameter, $s$~\cite{lanczos,fanchi,rohrlich,struck}.
The action functional then writes in terms of an
\emph{extended Lagrangian} $L_{\e}$
\begin{align}
S_{\e}[\bq(s),t(s)]&=\int_{s_{a}}^{s_{b}}L_{\e}\left(\bq,\dfrac{\bq}{s},t,
\dfrac{t}{s}\right)\d s\nonumber\\
&=\int_{s_{a}}^{s_{b}}L_{\e}\left(q^{\mu},\dfrac{q^{\mu}}{s}\right)\d s.
\label{principle1}
\end{align}
Herein, the index $\mu=0,\ldots,n$ denotes the entire range of
extended configuration space variables.
We consider extended Lagrangians with no explicit dependence on $s$.
As the action functional~(\ref{principle1}) has the form of
(\ref{principle0}), the subsequent Euler-Lagrange equations that determine
the particular path on which the value of the
functional~(\ref{principle1}) is stationary for variations with fixed endpoints, adopt the customary form
\begin{equation}\label{lageqm}
\dfrac{}{s}\left(\pfrac{L_{\e}}{\left(\dfrac{q^{\mu}}{s}
\right)}\right)-\pfrac{L_{\e}}{q^{\mu}}=0.
\end{equation}
For the index $\mu=0$, the Euler-Lagrange equation can be
expressed equivalently in terms of $t(s)$ as
\begin{equation}\label{lageqm-t}
\dfrac{}{s}\left(\pfrac{L_{\e}}{\left(\dfrac{t}{s}
\right)}\right)-\pfrac{L_{\e}}{t}=0.
\end{equation}
The equations of motion for both $\bq(s)$ and $t(s)$
are thus determined by the extended Lagrangian $L_{\e}$.
For a direct reparametrization of the conventional action, we take $\d t/\d s>0$ and denote the resulting extended Lagrangian by $L_{\e}^{\mathrm{h}}$.
Equality of the action functionals for arbitrary parametrized paths is then obtained by setting
$$
\int_{s_{a}}^{s_{b}}L\dfrac{t}{s}\,\d s=
\int_{s_{a}}^{s_{b}}L_{\e}^{\mathrm{h}}\,\d s.
$$
The extended Lagrangian $L_{\e}^{\mathrm{h}}$ is thus related to
the conventional Lagrangian, $L$, by
\begin{equation}\label{lag1}
L_{\e}^{\mathrm{h}}\left(\bq,\dfrac{\bq}{s},t,\dfrac{t}{s}\right)=
L\left(\bq,\dfrac{\bq}{t},t\right)\dfrac{t}{s},\qquad
\dfrac{\bq}{t}=\dfrac{\bq/\d s}{t/\d s}.
\end{equation}
The derivatives of $L_{\e}^{\mathrm{h}}$ from Eq.~(\ref{lag1}) with respect
to its arguments can now be expressed in terms of the conventional
Lagrangian $L$ as
\begin{align}
\quad\pfrac{L_{\e}^{\mathrm{h}}}{q^{\mu}}&=\pfrac{L}{q^{\mu}}\dfrac{t}{s},
\qquad\mu=1,\ldots,n\\
\quad\pfrac{L_{\e}^{\mathrm{h}}}{t}&=\pfrac{L}{t}
\dfrac{t}{s}\\
\pfrac{L_{\e}^{\mathrm{h}}}{\left(\dfrac{q^{\mu}}{s}\right)}&=
\pfrac{L}{\left(\dfrac{q^{\mu}}{t}\right)},\qquad
\mu=1,\ldots,n\label{L1-deri}\\
\pfrac{L_{\e}^{\mathrm{h}}}{\left(\dfrac{t}{s}\right)}&=L+
\sum_{\mu=1}^{n}\pfrac{L}{\left(\dfrac{q^{\mu}}{t}\right)}
\pfrac{\left(\dfrac{q^{\mu}/\d s}{t/\d s}\right)}{\left(\dfrac{t}{s}\right)}\dfrac{t}{s}
\nonumber\\
&=L-\sum_{\mu=1}^{n}\pfrac{L}{\left(\dfrac{q^{\mu}}{t}\right)}
\dfrac{q^{\mu}}{t}.\label{L1-deri2}
\end{align}
Equations~(\ref{L1-deri}) and (\ref{L1-deri2}) can be combined to yield the
following sum over the extended range $\mu=0,\ldots,n$ of
dynamical variables
\begin{align*}
\sum_{\mu=0}^{n}\pfrac{L_{\e}^{\mathrm{h}}}{\left(\dfrac{q^{\mu}}{s}\right)}
\dfrac{q^{\mu}}{s}&\equiv L_{\e}^{\mathrm{h}}.
\end{align*}
An extended Lagrangian $L_{\e}^{\mathrm{h}}$ can thus be set up immediately
by multiplying a given conventional Lagrangian $L$ with $\d t/\d s$
and expressing all velocities $\d\bq/\d t$ in terms of $\d\bq/\d s$
according to Eq.~(\ref{lag1}).
Such an extended Lagrangian $L_{\e}^{\mathrm{h}}$ may be referred to as a \emph{trivial extended Lagrangian}
since it contains no additional information on the underlying dynamical system.
A trivial extended Lagrangian $L_{\e}^{\mathrm{h}}$ constitutes a \emph{homogeneous function of degree one}
in the $n+1$ variables $\d q^{0}/\d s,\ldots,\d q^{n}/\d s$.
This may be seen by replacing all derivatives $\d q^{\mu}/\d s$
with $a\cdot\d q^{\mu}/\d s$, $a>0$ in Eq.~(\ref{lag1}), which yields
$$
L_{\e}^{\mathrm{h}}\left(\bq,a\dfrac{\bq}{s},t,a\dfrac{t}{s}\right)=
aL_{\e}^{\mathrm{h}}\left(\bq,\dfrac{\bq}{s},t,\dfrac{t}{s}\right).
$$
The preceding sum is therefore Euler's identity for a homogeneous function\cite{lanczos}.
For $L_{\e}^{\mathrm{h}}$, the time equation~(\ref{lageqm-t}) is a consequence of the spatial Euler-Lagrange equations and does not determine $t(s)$ independently.
The parametrization of time $t(s)$ is thus left undetermined ---
which reflects the fact that a conventional Lagrangian does not
provide any information on a parametrization of time and that
a trivial extended Lagrangian does not incorporate additional information.

A different construction is possible when a regular extended Lagrangian $L_{\e}$, which is not homogeneous of degree one in the velocities, reproduces the conventional dynamics on a selected invariant hypersurface.
This agreement on the hypersurface must be established separately, as will be done for the particle example below.
It does not extend the unrestricted identity~(\ref{lag1}) to a non-homogeneous Lagrangian.
In particular, the independent velocity derivatives in Eqs.~(\ref{L1-deri}) and (\ref{L1-deri2}) cannot be inferred from equality restricted to a hypersurface.

Regularity means that the velocity Hessian is nonsingular,
$$
\det\left[\pfrac{^{2}L_{\e}}
{\left(\dfrac{q^{\mu}}{s}\right)\partial\left(\dfrac{q^{\nu}}{s}\right)}\right]\ne0.
$$
It ensures that the extended Legendre transformation is locally invertible and that the Euler-Lagrange equations~(\ref{lageqm}) determine all accelerations.
Non-homogeneity alone does not ensure this property\cite{dirac,goldstein,johns}.
For the physical representation considered below, the selected hypersurface is defined by
\begin{equation}\label{lagid}
L_{\e}-\sum_{\mu=0}^{n}\pfrac{L_{\e}}{\left(\dfrac{q^{\mu}}{s}\right)}
\dfrac{q^{\mu}}{s}=-H_{\e}=0,
\end{equation}
where $H_{\e}$ is the extended energy, or equivalently the extended Hamiltonian after the Legendre transformation.
Equation~(\ref{lagid}) is imposed as a condition on the initial data.
Its preservation follows by calculating the total $s$-derivative of its left-hand side and inserting the Euler-Lagrange equations~(\ref{lageqm})
\begin{align}
&\quad\,\dfrac{}{s}L_{\e}\left(q^{\mu},\dfrac{q^{\mu}}{s}\right)-
\sum_{\mu=0}^{n}\dfrac{q^{\mu}}{s}\dfrac{}{s}\pfrac{L_{\e}}
{\left(\dfrac{q^{\mu}}{s}\right)}\nonumber\\
&\qquad-\sum_{\mu=0}^{n}\pfrac{L_{\e}}
{\left(\dfrac{q^{\mu}}{s}\right)}\dfrac{\left(\dfrac{q^{\mu}}{s}\right)}{s}\nonumber\\
&=\dfrac{L_{\e}}{s}-\sum_{\mu=0}^{n}\pfrac{L_{\e}}{q^{\mu}}\dfrac{q^{\mu}}{s}-\sum_{\mu=0}^{n}
\pfrac{L_{\e}}{\left(\dfrac{q^{\mu}}{s}\right)}\dfrac{\left(\dfrac{q^{\mu}}{s}\right)}{s}\nonumber\\
&=0.
\label{lagid-deri2}
\end{align}
Equation~(\ref{lagid-deri2}) establishes conservation of $H_{\e}$.
Its value is determined by the initial data, and the zero value in Eq.~(\ref{lagid}) is an additional physical selection.
This is analogous to the case of a conventional Hamiltonian system with no \emph{explicit} time dependence,
$H(\bq,\bp)=\Hv_{0}$, where the system's initial energy $\Hv_{0}$ embodies a \emph{constant of motion}.
The equations of motion preserve the selected energy level.
In the language of Differential Geometry, the system's motion then takes place on a \emph{hypersurface}
that is then defined by $H(\bq,\bp)=\Hv_{0}$ within the cotangent
bundle $T^{*}\MB$ over the configuration manifold $\MB$.

In the actual case, the selected physical motion along $s$ takes place on a hypersurface
that is defined by Eq.~(\ref{lagid}), which resides in the extended tangent bundle $T(\MB\times\RB)$ over
the space-time configuration manifold $\MB\times\RB$.
This contrasts with the conventional Lagrangian
description occurring in $(T\MB)\times\RB$.
The regular Legendre transformation maps the selected hypersurface into the symplectic extended phase space $T^{*}(\MB\times\RB)$ of dimension $2n+2$.
The ambient phase space is symplectic, whereas the restricted two-form on a regular energy hypersurface of dimension $2n+1$ is degenerate.

An example of a non-trivial (non-homogeneous) extended Lagrangian
is furnished by the following $L_{\e}$ that describes, on its zero-energy hypersurface, the motion of a relativistic spinless
point particle of mass $m>0$ and charge $\zeta$ in an external electromagnetic
field defined by the potentials $(\phi,\bA)$
\begin{align}
L_{\e}\left(\bq,\dfrac{\bq}{s},t,\dfrac{t}{s}\right)&=
\onehalf mc^{2}\left[{\frac{1}{c^{2}}\left(\dfrac{\bq}{s}\right)}^{2}-
{\left(\dfrac{t}{s}\right)}^{2}-1\right]\nonumber\\
&\mbox{}\quad+\frac{\zeta}{c}\bA(\bq,t)
\dfrac{\bq}{s}-\zeta\,\phi(\bq,t)\dfrac{t}{s}.
\label{lag1-em}
\end{align}
The associated hypersurface condition~(\ref{lagid}) for $L_{\e}$ coincides with that
for the free-particle Lagrangian as all terms linear in the velocities drop out
\begin{equation}\label{hypersurface-em}
{\left(\dfrac{t}{s}\right)}^{2}-\frac{1}{c^{2}}
{\left(\dfrac{\bq}{s}\right)}^{2}=1\;\;\Leftrightarrow\;\;
\dfrac{s}{t}=\sqrt{1-\frac{1}{c^{2}}{\left(\dfrac{\bq}{t}\right)}^{2}}.
\end{equation}
Here the future-directed branch $\d t/\d s>0$ is understood.
The selected hypersurface~(\ref{hypersurface-em}) identifies $s$ with the particle's proper time, up to an additive constant, and fixes its correlation with laboratory time.
With Eq.~(\ref{hypersurface-em}), the extended
Lagrangian~(\ref{lag1-em}) may be projected into $(T\MB)\times\RB$
by setting $L=L_{\e}\,\d s/\d t$ on this hypersurface
to yield the well-known conventional relativistic Lagrangian $L$ of the actual physical system
\begin{equation}\label{lagr-em}
L\left(\bq,\dfrac{\bq}{t},t\right)=
-mc^{2}\sqrt{1-\frac{1}{c^{2}}{\left(
\dfrac{\bq}{t}\right)}^{2}}+\frac{\zeta}{c}\bA
\dfrac{\bq}{t}-\zeta\,\phi.
\end{equation}
The quadratic form of the velocity terms in the Lagrangian~(\ref{lag1-em}) is lost owing to the projection.
The square-root dependence in Eq.~(\ref{lagr-em}) prevents the elementary Gaussian evaluation used below.
The equivalence of the classical equations on the selected hypersurface will be verified explicitly, and the quantum correspondence will be established by deriving the physical wave equation.

For the following derivations, it is helpful to rewrite the Lagrangian~(\ref{lag1-em})
in covariant notation.
With Einstein's summation convention and the notation
$A_{0}(q^{\mu})=-\phi(q^{\mu})$ for the metric
$\eta_{\mu\nu}=\mathrm{diag}(-1,1,1,1)=\eta^{\mu\nu}$,
the nonhomogeneous extended Lagrangian~(\ref{lag1-em}) writes
\begin{equation}\label{lag1-em2}
L_{\e}\left(q^{\mu},\dfrac{q^{\mu}}{s}\right)=
\onehalf m\dfrac{q^{\alpha}}{s}\dfrac{q_{\alpha}}{s}+
\frac{\zeta}{c}A_{\alpha}\dfrac{q^{\alpha}}{s}-\onehalf mc^{2}.
\end{equation}
The hypersurface condition~(\ref{hypersurface-em}) is then converted into
\begin{equation}\label{hypersurface-em2}
\dfrac{q^{\alpha}}{s}\dfrac{q_{\alpha}}{s}=-c^{2}.
\end{equation}
The velocity Hessian of Eq.~(\ref{lag1-em2}) is $m\eta_{\alpha\beta}$ and is nonsingular for $m>0$.
Its canonical momenta and extended Hamiltonian are
\begin{align*}
p_{\alpha}&=m\dfrac{q_{\alpha}}{s}+\frac{\zeta}{c}A_{\alpha},\\
H_{\e}&=\onehalf m\dfrac{q^{\alpha}}{s}\dfrac{q_{\alpha}}{s}+\onehalf mc^{2}\\
&=\frac{1}{2m}\left(p_{\alpha}-\frac{\zeta}{c}A_{\alpha}\right)
\left(p^{\alpha}-\frac{\zeta}{c}A^{\alpha}\right)+\onehalf mc^{2}.
\end{align*}
The selected condition $H_{\e}=0$ is therefore the usual electromagnetic mass shell.
The Euler-Lagrange equations take the Lorentz-force form
$$
m\dfrac{}{s}\left(\dfrac{q_{\alpha}}{s}\right)
=\frac{\zeta}{c}F_{\alpha\beta}\dfrac{q^{\beta}}{s},
\qquad
F_{\alpha\beta}=\pfrac{A_{\beta}}{q^{\alpha}}-\pfrac{A_{\alpha}}{q^{\beta}}.
$$
Antisymmetry of $F_{\alpha\beta}$ implies conservation of the squared four-velocity, so the normalization~(\ref{hypersurface-em2}) is preserved once it is selected.

For timelike velocities, the homogeneous extension of Eq.~(\ref{lagr-em}) is
$$
L_{\e}^{\mathrm{h}}
=-mc\sqrt{-\dfrac{q^{\alpha}}{s}\dfrac{q_{\alpha}}{s}}
+\frac{\zeta}{c}A_{\alpha}\dfrac{q^{\alpha}}{s}.
$$
Its difference from Eq.~(\ref{lag1-em2}) satisfies
$$
L_{\e}-L_{\e}^{\mathrm{h}}
=-\frac{m}{2}\left(\sqrt{-\dfrac{q^{\alpha}}{s}\dfrac{q_{\alpha}}{s}}-c\right)^{2}.
$$
Consequently, the two Lagrangians and their first coordinate and velocity derivatives coincide on the hypersurface~(\ref{hypersurface-em2}).
Their Euler-Lagrange equations agree along curves remaining on this hypersurface.
The non-homogeneous extended Lagrangian~(\ref{lag1-em2}) therefore describes the same physical classical motion as Eq.~(\ref{lagr-em}) when the proper-time normalization is selected.
The conventional action $L\,\d t$ is Lorentz invariant.

In Feynman's famous path integral approach to non-relativistic quantum mechanics\cite{feynman},
the space and time evolution of a wave function $\psi(\bq,t)$
is formulated in terms of a transition amplitude density
$K(b,a)$, also referred to as a \emph{kernel}, or, a \emph{propagator}.
To treat space and time on equal footing, we use the extended Lagrangian description and introduce a wave function $\psi(q^{\mu},s)$.
In the path integral, $s$ is an auxiliary worldline parameter that coincides with proper time on the normalized timelike classical trajectories.
For an infinitesimal positive step $\epsilon=s_{b}-s_{a}$, we may approximate
the action functional $S_{\e}$ from Eq.~(\ref{principle1}) by
$$
S_{\e,\epsilon}[q^{\mu}(s)]=\epsilon\,L_{\e}\left(\frac{q^{\mu}_{b}+
q^{\mu}_{a}}{2},\frac{q^{\mu}_{b}-q^{\mu}_{a}}{\epsilon}\right).
$$
The transition of a given wave function $\psi(q^{\mu}_{a},s_{a})$
at the parameter value $s_{a}$ to the wave function
$\psi(q^{\mu}_{b},s_{a}+\epsilon)$ at $s_{b}=s_{a}+\epsilon$ can now
be formulated as the path integral
\begin{equation}\label{trans-infini}
\psi(q^{\mu}_{b},s_{a}+\epsilon)=\frac{1}{M}\int\exp\left[
\frac{i}{\hbar}S_{\e,\epsilon}\right]\psi(q^{\mu}_{a},s_{a})\,\d^{4}q_{a}.
\end{equation}
Note that we integrate here over the entire space-time.
The integration includes histories that do not obey the classical normalization~(\ref{hypersurface-em2}) at every step.
The physical mass shell will be imposed as a condition on quantum states after deriving their auxiliary evolution.
The yet to be determined normalization factor
$M$ represents the integration measure for the
``infinitesimal step'' path integral~(\ref{trans-infini}).
Clearly, this measure depends on the step size~$\epsilon$.
To test this approach, we derive the auxiliary evolution equation and then the Klein-Gordon equation on the basis of
the extended Lagrangian $L_{\e}$ for a relativistic point particle
in an external electromagnetic field from Eq.~(\ref{lag1-em2}).

For an infinitesimal parameter step $\epsilon\equiv\Delta s$,
the action $S_{\e,\epsilon}$ for the extended
Lagrangian~(\ref{lag1-em2}) writes
\begin{align}
\qquad S_{\e,\epsilon}=\epsilon L_{\e}&=\onehalf m
\frac{\eta_{\alpha\beta}(q^{\alpha}_{b}-q^{\alpha}_{a})
(q^{\beta}_{b}-q^{\beta}_{a})}{\epsilon}\nonumber\\
&\mbox{}\quad+\frac{\zeta}{c}(q^{\alpha}_{b}-q^{\alpha}_{a})\,
A_{\alpha}(q^{\mu}_{c})-\onehalf mc^{2}\epsilon.
\label{action1}
\end{align}
The potentials $A_{\alpha}$ are to be taken at the space-time
location $q^{\mu}_{c}=(q^{\mu}_{b}+q^{\mu}_{a})/2$.
We insert this particular action function into Eq.~(\ref{trans-infini})
and perform a transformation of the integration variables $q^{\mu}_{a}$,
$$
q^{\mu}_{b}-q^{\mu}_{a}=\xi^{\mu}\quad\Rightarrow\quad
\d^{4}q_{a}=\d^{4}\xi.
$$
The integral~(\ref{trans-infini}) has now the equivalent representation
\begin{equation}\label{trans-infini2}
\psi(q^{\mu}_{b},s_{a}+\epsilon)=\frac{1}{M}\int\exp\left[
\frac{i}{\hbar}S_{\e,\epsilon}\right]\psi(q^{\mu}_{b}-\xi^{\mu},s_{a})\,\d^{4}\xi,
\end{equation}
while the action $S_{\e,\epsilon}$ from Eq.~(\ref{action1}) takes on the form
$$
S_{\e,\epsilon}=\frac{m}{2}\frac{\eta_{\alpha\beta}
\xi^{\alpha}\xi^{\beta}}{\epsilon}+
\frac{\zeta}{c}\xi^{\alpha}\!\left[A_{\alpha}(q^{\mu}_{b})\!-\!\onehalf
\xi^{\beta}\pfrac{A_{\alpha}(q^{\mu}_{b})}{q^{\beta}}\right]-
\epsilon\frac{mc^{2}}{2}.
$$
Herein, the potentials $A_{\alpha}(q^{\mu}_{c})$ were expressed
to first order in terms of their values at $q^{\mu}_{b}$.
In the following, we skip the index ``$b$'' in the coordinate vector
since all $q^{\mu}$ refer to that particular space-time event
from this point of our derivation.
We also write $s=s_{a}$.

In order to match the quadratic terms in $S_{\e,\epsilon}$,
the wave function $\psi(q^{\mu}-\xi^{\mu},s)$ under the
integral~(\ref{trans-infini2}) must be expanded up to
second order in the $\xi^{\mu}$,
\begin{align*}
\psi(q^{\mu}-\xi^{\mu},s)&=\psi(q^{\mu},s)-\xi^{\alpha}
\pfrac{\psi(q^{\mu},s)}{q^{\alpha}}\\
&\quad+\onehalf\xi^{\alpha}\xi^{\beta}
\pfrac{^{2}\psi(q^{\mu},s)}{q^{\alpha}\partial q^{\beta}}+\mathcal O(\xi^{3}).
\end{align*}
In the following expressions, unmarked $\psi$ and its derivatives on the right-hand side are evaluated at $(q^{\mu},s)$.
The rest energy term in $S_{\e,\epsilon}$ depends only on $\epsilon$.
It can, therefore, be taken as a factor in front of the integral
and expanded up to first order in $\epsilon$.
The total expression~(\ref{trans-infini2}) for the transition
of the wave function $\psi$ thus follows as
\begin{widetext}
\begin{equation}\label{trans-infini3}
\begin{aligned}
\psi(q^{\mu},s+\epsilon)=\frac{1}{M}
\left(1-\epsilon\frac{imc^{2}}{2\hbar}\right)
&\int_{-\infty}^{\infty}\exp\left[\frac{i}{\hbar\epsilon}\left(
\frac{m}{2}\eta_{\alpha\beta}\xi^{\alpha}\xi^{\beta}+
\frac{\zeta\epsilon}{c}A_{\alpha}\xi^{\alpha}-\frac{\zeta\epsilon}{2c}
\pfrac{A_{\alpha}}{q^{\beta}}\xi^{\alpha}\xi^{\beta}\right)\right]\\
&\quad\times\left(\psi-\xi^{\alpha}
\pfrac{\psi}{q^{\alpha}}+\onehalf\xi^{\alpha}\xi^{\beta}
\pfrac{^{2}\psi}{q^{\alpha}\partial q^{\beta}}\right)\d^{4}\xi.
\end{aligned}
\end{equation}
Prior to calculating the Gaussian type integrals, we specify them as Fresnel limits over real integration variables.
A factor $\exp[-\delta\sum_{\mu=0}^{3}(\xi^{\mu})^{2}]$, with $\delta>0$, is inserted temporarily and the limit $\delta\to0^{+}$ is taken after integration.
In the short-time expansion, quadratic insertions in $\xi^{\mu}$ contribute at order $\epsilon$.
The potential-gradient term in the exponent of Eq.~(\ref{trans-infini3}) can therefore be expanded as
$$
\exp\left(-\frac{i\zeta}{2\hbar c}\pfrac{A_{\alpha}}{q^{\beta}}
\xi^{\alpha}\xi^{\beta}\right)=1-\frac{i\zeta}{2\hbar c}
\pfrac{A_{\alpha}}{q^{\beta}}\xi^{\alpha}\xi^{\beta}+\ldots
$$
Omitting terms of higher order than quadratic in the $\xi^{\mu}$,
the integral becomes
\begin{align*}
\psi(q^{\mu},s+\epsilon)=\frac{1}{M}
\left(1-\epsilon\frac{imc^{2}}{2\hbar}\right)
&\int_{-\infty}^{\infty}\exp\left[\frac{i}{\hbar}\left(
\frac{m}{2\epsilon}\eta_{\alpha\beta}\xi^{\alpha}\xi^{\beta}+
\frac{\zeta}{c}A_{\alpha}\xi^{\alpha}\right)\right]\\
&\quad\times\left[\psi-\xi^{\alpha}
\pfrac{\psi}{q^{\alpha}}+\onehalf\xi^{\alpha}\xi^{\beta}\left(
\pfrac{^{2}\psi}{q^{\alpha}\partial q^{\beta}}-\frac{i\zeta}{\hbar c}
\pfrac{A_{\alpha}}{q^{\beta}}\,\psi\right)\right]\d^{4}\xi.
\end{align*}
The integral over the entire space-time can now be solved
analytically to yield
\begin{align*}
\psi(q^{\mu},s+\epsilon)&=\frac{i}{M}{\left(
\frac{2\pi\hbar\epsilon}{m}\right)}^{2}
\left(1-\epsilon\frac{imc^{2}}{2\hbar}\right)
\exp\left(-\epsilon\frac{i\zeta^{2}}{2\hbar mc^{2}}
A^{\alpha}A_{\alpha}\right)\\
&\quad\times\left[\psi+\epsilon\frac{\zeta}{mc}A^{\alpha}
\pfrac{\psi}{q^{\alpha}}+\frac{\epsilon}{2}\left(
\pfrac{^{2}\psi}{q^{\alpha}\partial q^{\beta}}
-\frac{i\zeta}{\hbar c}\pfrac{A_{\alpha}}{q^{\beta}}\psi\right)
\left(\frac{\epsilon\zeta^{2}}{m^{2}c^{2}}A^{\alpha}A^{\beta}+
\frac{i\hbar}{m}\eta^{\alpha\beta}\right)\right].
\end{align*}
We may omit the term quadratic in $\epsilon$ that is contained
in the rightmost factor and finally expand the exponential function
up to first order in $\epsilon$
\begin{align*}
\psi(q^{\mu},s+\epsilon)&=\frac{i}{M}{\left(\frac{2\pi\hbar\epsilon}{m}\right)}^{2}
\left(1-\epsilon\frac{imc^{2}}{2\hbar}\right)
\left(1-\epsilon\frac{i\zeta^{2}}{2\hbar mc^{2}}
A^{\alpha}A_{\alpha}\right)\\
&\quad\!\times\left[\psi+\epsilon\frac{\zeta}{mc}A^{\alpha}
\pfrac{\psi}{q^{\alpha}}+\epsilon\frac{i\hbar}{2m}\left(
\pfrac{^{2}\psi}{q^{\alpha}\partial q_{\alpha}}
-\frac{i\zeta}{\hbar c}\pfrac{A^{\alpha}}{q^{\alpha}}\psi\right)\right]
\end{align*}
\end{widetext}
Normalizing the kinetic kernel gives
$$
M=i\left(\frac{2\pi\hbar\epsilon}{m}\right)^{2}.
$$
For the metric $\eta_{\mu\nu}$, the temporal Fresnel factor has phase $\exp(-i\pi/4)$ and the three spatial factors together have phase $\exp(3i\pi/4)$, giving the factor $i$ in $M$.
Expanding the left-hand side as
$$
\psi(q^{\mu},s+\epsilon)
=\psi(q^{\mu},s)+\epsilon\pfrac{\psi}{s}+\mathcal O(\epsilon^{2})
$$
and equating terms linear in $\epsilon$ yields
\begin{equation}\label{s-evolution}
i\hbar\pfrac{\psi}{s}=\widehat H_{\e}\psi,
\qquad
\widehat H_{\e}=-\frac{\hbar^{2}}{2m}D_{\alpha}D^{\alpha}+\frac{mc^{2}}{2},
\end{equation}
where
$$
D_{\alpha}=\pfrac{}{q^{\alpha}}-\frac{i\zeta}{\hbar c}A_{\alpha}.
$$
The operator $\widehat H_{\e}$ is the quantized extended Hamiltonian with the ordering fixed by the midpoint prescription.
In particular, this prescription supplies the term involving $\partial A^{\alpha}/\partial q^{\alpha}$ in the product $D_{\alpha}D^{\alpha}$.

We implement the classical selection $H_{\e}=0$ by imposing the quantum mass-shell condition
\begin{equation}\label{quantum-constraint}
\widehat H_{\e}\psi=0.
\end{equation}
As $\widehat H_{\e}$ has no explicit $s$ dependence, this condition is preserved by Eq.~(\ref{s-evolution}).
Equations~(\ref{s-evolution}) and (\ref{quantum-constraint}) imply $\partial\psi/\partial s=0$ in the physical sector, for which we write $\psi=\psi(q^{\mu})$.
In this sector the sum of the five terms linear in $\epsilon$ vanishes and establishes the equation
$$
\frac{m^{2}c^{2}}{\hbar^{2}}\psi=
\pfrac{^{2}\psi}{q^{\alpha}\partial q_{\alpha}}
-\frac{\zeta^{2}A^{\alpha}A_{\alpha}}{\hbar^{2}c^{2}}
\psi+\frac{2\zeta A^{\alpha}}{i\hbar c}
\pfrac{\psi}{q^{\alpha}}+\frac{\zeta}{i\hbar c}
\pfrac{A^{\alpha}}{q^{\alpha}}\psi.
$$
This equation has the equivalent product form
$$
\left(\pfrac{}{q^{\alpha}}-\frac{i\zeta}{\hbar c}A_{\alpha}\right)
\left(\pfrac{}{q_{\alpha}}-\frac{i\zeta}{\hbar c}A^{\alpha}\right)
\psi={\left(\frac{mc}{\hbar}\right)}^{2}\psi,
$$
which constitutes exactly the Klein-Gordon equation
for a charged spin-zero particle with our metric $\eta_{\mu\nu}$.
The kernel in Eq.~(\ref{trans-infini}) describes evolution in the auxiliary parameter $s$.
Composing these short-time kernels gives the kernel for a finite auxiliary interval.
The space-time Klein-Gordon Green function is obtained, up to the conventional overall normalization, by integrating the latter kernel over positive values of $s_{b}-s_{a}$ with the appropriate causal prescription\cite{feynman50}.

We remark that Feynman~\cite{feynman48} already sketched a derivation of the Klein-Gordon equation using a quadratic four-dimensional Lagrangian and a specified dependence of the wave function on an auxiliary parameter.
His description of those results as ``purely formal'' concerned their interpretation in that discussion.
In his later treatment\cite{feynman50}, Appendix~A explicitly separates auxiliary evolution from fixed-mass selection and constructs the Klein-Gordon propagator by a proper-time integral.
Up to conventions and a rescaling of the auxiliary parameter, the constant term $-\onehalf mc^{2}$ in Eq.~(\ref{lag1-em2}) incorporates the corresponding mass phase into the kernel.
The present derivation makes explicit the relation between this proper-time construction and the classical extended Lagrangian on its selected mass shell.

\emph{Conclusions}
Starting from the space-time formulation of the action principle,
it was demonstrated that the Lagrangian description of classical
dynamics can be reformulated in terms of extended Lagrangians
in order to put space and time on equal footing.
The direct reparametrization of a conventional action yields a homogeneous extended Lagrangian.
For a relativistic charged particle, a regular non-homogeneous extended Lagrangian reproduces the same classical dynamics on a selected zero-energy hypersurface.
The equations of motion preserve this mass shell, whose selection identifies the classical evolution parameter with proper time.
Due to the quadratic velocity dependence, the corresponding path integral yields an auxiliary evolution equation by Gaussian integration.
Imposing the quantum mass-shell condition then gives the minimally coupled Klein-Gordon equation.
This connects the extended classical formulation with the established proper-time approach to relativistic quantum physics.
\vspace*{-1mm}


\begin{thebibliography}{10}
\bibitem{lanczos}
C.~Lanczos, {\it The Variational Principles of Mechanics\/}
(University of Toronto Press, Toronto, Ontario, 1949),
Reprint 4th edn (Dover Publications, New York, 1986).
\bibitem{fanchi}
J.R.~Fanchi, {\it Parametrized Relativistic Quantum Theory\/}
(Kluwer Academic Publishers, Dordrecht, The Netherlands, 1993).
\bibitem{rohrlich}
F.~Rohrlich, Ann.~Phys.~(N.Y.)~{\bf 117}, 292 (1979).
\bibitem{struck}
J.~Struckmeier, J.~Phys.~A:~Math.~Gen.~{\bf 38}, 1257 (2005).
\bibitem{dirac}
P.A.M.~Dirac, Can.~J.~Math.~{\bf 2}, 129--148 (1950).
\bibitem{goldstein} H.~Goldstein, C.~Poole, and J.~Safko,
\emph{Classical Mechanics}, 3rd ed.\
(Pearson, Addison-Wesley, Upper Saddle River, NJ, 2002).
\bibitem{johns}
O.D.~Johns, {\it Analytical Mechanics for Relativity and
Quantum Mechanics\/} (Oxford University Press, Oxford, 2005).
\bibitem{feynman}
R.P.~Feynman and A.R.~Hibbs, {\it Quantum Mechanics and Path Integrals\/}
(Emended Edition by Daniel F.~Styer, Dover Publications, Inc., Mineola, New York, 2010).
\bibitem{feynman50}
R.P.~Feynman, Phys.~Rev.~{\bf 80}, 440--457 (1950).
\bibitem{feynman48}
R.P.~Feynman, Rev.~Mod.~Phys.~{\bf 20}, 367 (1948).
\end{thebibliography}
\end{document}